ا
\documentclass[
 aip,
 jmp,%
 amsmath,amssymb,
superscriptaddress,
reprint,
]{revtex4-1}
\usepackage{comment}
\usepackage{graphicx}
\usepackage{dcolumn}
\usepackage{bm}
\usepackage{hyperref}
\usepackage{xcolor}
\hypersetup{
    colorlinks=true,
    linkcolor=blue,
    citecolor=blue,
    urlcolor=cyan,
}

\begin{document}

\preprint{AIP/123-QED}

\title{Width dependent auto-oscillating properties of constriction based spin Hall nano-oscillators.}
\author{Ahmad A. Awad}
\author{Afshin Houshang}
\author{Mohammad Zahedinejad}
\affiliation{Department of Physics, University of Gothenburg, 412 96 Gothenburg, Sweden}
\author{Roman Khymyn}
\affiliation{Department of Physics, University of Gothenburg, 412 96 Gothenburg, Sweden}
\author{Johan \AA kerman}
\affiliation{Department of Physics, University of Gothenburg, 412 96 Gothenburg, Sweden}
\affiliation{Materials and Nanophysics, School of Engineering Sciences, KTH Royal Institute of Technology, 164 00 Kista, Sweden}

\date{\today}

\begin{abstract}
We study the current tunable microwave signal properties of nano-constriction based spin Hall nano-oscillators (SHNOs) in oblique magnetic fields as a function of the nano-constriction width, $w=$~50--140 nm. The threshold current is found to scale linearly with $w$, defining a constant threshold current density of $J_{th}=$ 1.7 $\times$ 10$^{8}$ A/cm$^2$. While the current dependence of the microwave frequency shows the same generic non-monotonic behavior for all $w\geqslant$ 80 nm, the quality of the generated microwave signal improves strongly with $w$, showing a linear $w$ dependence for both the total power and the linewidth. 
As a consequence, the peak power for a 140 nm nano-constriction is about an order of magnitude higher than that of a 80 nm nano-constriction. The smallest nano-constriction, $w=$ 50 nm, exhibits a different behavior with a higher power and a worse linewidth indicating a crossover into a qualitatively different narrow-constriction regime. 
\end{abstract}


\maketitle

Spin Hall nano-oscillators (SHNOs)
are a class of miniaturized, ultra-broadband, microwave signal generators~\cite{demidov2012ntm,liu2012prl, liu2013prl,demidov2014ntc,ulrichs2014apl,duan2014ntc, demidov2014apl,ranjbar2014ieeeml,Madami2015JAP,Mazraati2016apl,Chen2016procieee,divinskiy2017apl,demidov2017pr,Zahedinejad2018APL,mazraati2018pra,Sato2019PRL,Tarequzzaman2019CommPhys,divinskiy2019apl,chen2019pra,chen2019prb} where the spin Hall effect~\cite{hirsch1999prl, hoffmann2013ieeem, sinova2015rmp} converts a direct charge current into a transverse pure spin current, transferring angular momentum~\cite{slonczewski1996jmmm,berger1996prb,ralph2008jmmm} into a nearby ferromagnetic layer, resulting in current and magnetic field tunable spin wave (SW) auto-oscillations (AOs); a microwave voltage is then generated via the anisotropic magnetoresistance of the magnetic layer. Thanks to their simple bilayer structure, nano-constriction SHNOs offer a number of advantages such as straightforward fabrication processes, flexibility in material choices, significant freedom in layout, and direct optical access to the active auto-oscillating region, which makes them attractive for fundamental studies of highly nonlinear and interacting SWs. 

Depending on the magnetic material properties, the drive current, and the direction and magnitude of the magnetic field, a variety of nonlinear SW modes can be generated: 
localized edge modes~\cite{demidov2014apl,Awad2016NatPhys,Durrenfeld2017Nanosc}, magnetic droplets~\cite{divinskiy2017prb},and propagating SWs~\cite{Fulara2019SciAdv,divinskiy2018advmat}. For applications, nano-constrictions offer wide frequency tunability~\cite{Zahedinejad2018APL}, fast modulation rates~\cite{Zahedinejad2017IEEE}, easy injection locking~\cite{Hache2019APL}, and robust mutual synchronization in both one~\cite{Awad2016NatPhys}, and two dimensional networks~\cite{Zahedinejad2019NatN}. These properties make them ideal candidates for broadband microwave signal generation and processing, and non-conventional oscillator computing. 

Despite a number of recent nano-constriction studies, a systematic investigation of how their signal properties depend on the nano-constriciton width ($w$) is still lacking. In order to take advantage of their full potential, one needs to better understand how $w$ controls their threshold current, frequency, tunability, linewidth, and output power. 

Here we study the microwave signal generated in nano-constriction SHNOs with different constriction widths (50--140 nm) in oblique magnetic fields. The threshold current scales linearly with $w$, consistent with a single threshold current density independent of $w$. The frequency vs.~current dependence is found to be non-monotonic and exhibits the same generic dependence for nanoconstrictions with different widths, except for the most narrow constriction, which shows a much stronger blue-shifting behavior. Both the output power and the linewidth improve linearly with $w$, except for the 50 nm constriction which shows the highest output power but an order of magnitude worse linewidth. Our results indicate that \emph{i}) for larger constriction widths, $w\geqslant$ 80 nm, the SHNO behavior is consistent with a model \cite{Dvornik2018prappl} where the AO starts as an edge mode and expands into the interior of the constriction at higher currents, and \emph{ii}) there is a cross-over into a qualitatively different interior-only AO for $w=$ 50 nm, similar to what was recently observed for ultra-narrow (20 nm) constrictions \cite{Durrenfeld2017Nanosc}. %
The nano-constriction SHNO is fabricated from a bilayer of 5~nm Pt and 5~nm Py (Ni$_{80}$Fe$_{20}$), magnetron sputtered at room temperature onto a 20$\times$20~mm$^2$ piece of $c$-plane sapphire substrate in a system with a base pressure lower than 3$\times$10$^{-8}$~Torr. Before breaking vacuum, the bilayer is capped with 5~nm layer of SiO$_2$ to prevent oxidation. 
The bilayer was then patterned first by electron beam lithography of a negative resist into 4~$\mu$m$\times$12~$\mu$m rectangles with bow tie shaped nano-constrictions and then argon ion milling was used to transfer the pattern using the negative resist as the etching mask. A coplanar waveguide (CPW) provides electrical contacts is defined by optical lithography, followed by reactive ion etching of the protective SiO$_2$ layer, sputtering of copper, and lift-off. A scanning electron microscopy (SEM) image of the device is shown in Fig.1(a) together with a schematic layout. $w$ shows the width of the device ranging from 50 to 140 nm.
The lower inset in Fig. 1(a) shows the resistance of the SHNOs as a function of 1/$w$. The series resistance is calculated by finding the intercept of a linear fit (red line) with the $y$-axis, which is about 73 $\Omega$. $\theta$ shows the out-of-plane and $\varphi$ the in-plane direction of the external magnetic field, $H$.
Electrical measurements were done using a custom-built setup in which out-of-plane magnetic fields can be applied to the sample. A direct current was applied to the SHNO through a bias-T, the generated microwave signals were picked up via the high-frequency port, amplified by a low noise amplifier (LNA), and subsequently recorded by a spectrum analyzer. The recorded spectra were then processed to account for both the RF circuit gain from the LNA and losses from the impedance mismatch between the measurement line resistance (50 $\Omega$), and the total resistance of the CPW and the device. The resulting spectra is then fitted to a Lorentzian function to extract the power and linewidth.  Fig.1(b) demonstrates a schematic of the measurement setup.

\begin{figure}[t]
\centering
\includegraphics[width=6cm]{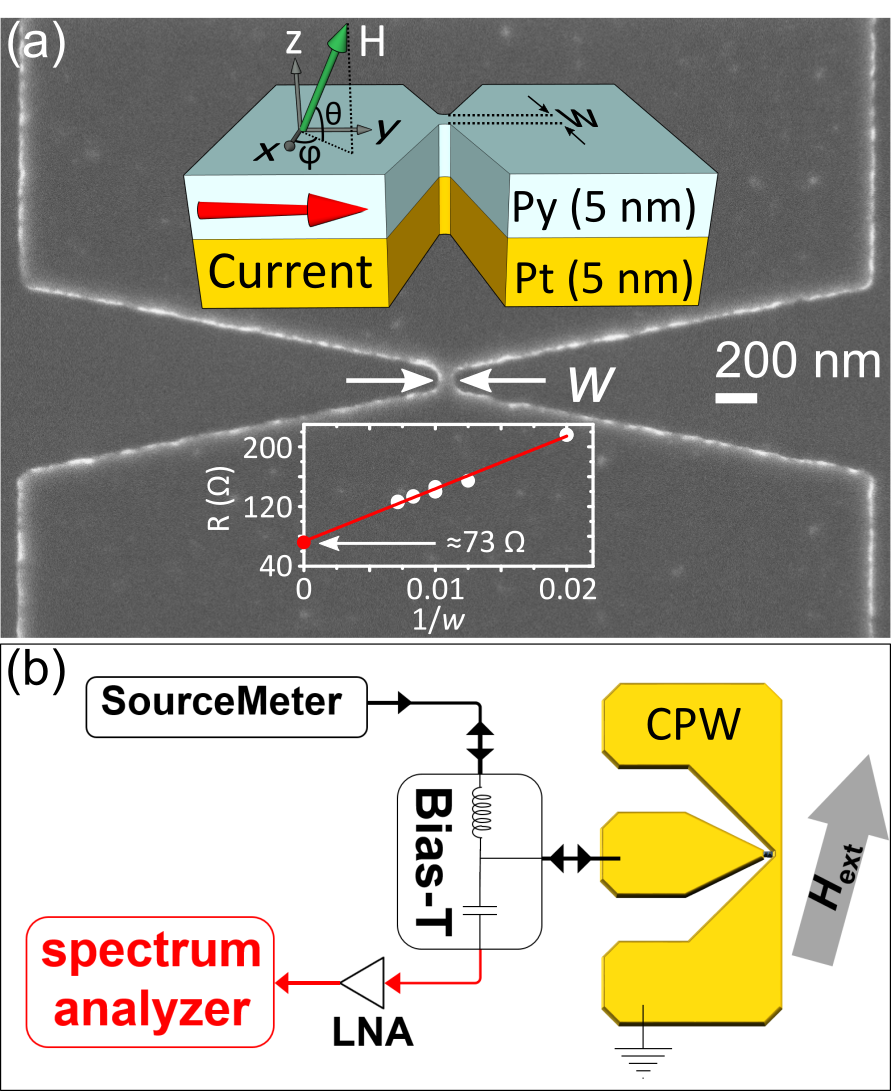}
\caption{(a) SEM image of a Py/Pt SHNO. $w$ is the width of the constriction. The upper inset shows the device schematic together with the applied field direction. The bottom inset shows the device resistance as a function of 1/$w$. The series resistance of the device leads is about 73 $\Omega$. (b) Schematic of the measurement setup.}
\label{fig:1} 
\end{figure}

To compare the magnetodynamical behaviour of SHNOs with different constriction sizes, the magnitude, out-of-plane, and in-plane angles of the applied magnetic field are kept exactly the same in all measurements, namely $\mu_{0}H=0.76$ T, $\theta=$ 80$^{\circ}$, and $\varphi=$ 22$^\circ$. The drive current is then varied accordingly based on the size of each device and the microwave emission is recorded. 
Fig.~2(a) shows the power spectral density of the emitted microwave response of a SHNO with $w=$ 120 nm. Fig.~2(b) and (c) show the integrated power and linewidth of the same device. The AO frequency, as can be seen in Fig.~2(a), shows a non-monotonic current dependence i.e.~red shifting at low currents and then blue shifting above a specific current. 

The behavior is conveniently described by the magnetodynamic nonlinearity, $\mathcal{N}$ \cite{slavin2009ieeem, Dvornik2018prappl}, with its sign determining the nature of magnon interactions within the system. A negative $\mathcal{N}$ makes the AO frequency decrease with amplitude, while a positive $\mathcal{N}$ makes it increase as:
\begin{equation}
    f_{(I_{dc})}= f_{0} +\frac{\mathcal{N}}{2\pi}P,
\end{equation}
where $P$ is the normalized AO power and $f_{0}$ is the AO frequency at threshold ($P\to 0$). Although $\mathcal{N}$ in principle depends on the full demagnetizing tensor, for wide enough constrictions one can use the thin film approximation \cite{gerhart2007prb}, where 
\begin{equation}
    \mathcal{N}=\frac{\omega_{M}\omega_{H}}{\omega_0}\left(\frac{3\omega_H^2 sin^2\theta_{int}}{\omega_0^2}-1\right),
\end{equation}
in which $\omega_0$ is the FMR frequency in the magnetic layer, $\omega_H=\gamma H_{int}$, and $\omega_M= \gamma M_0$; $H_{int}$ and $\theta_{int}$ are the internal field magnitdue and angle, respectively.

\begin{figure}[t]
\centering
\includegraphics[width=7cm]{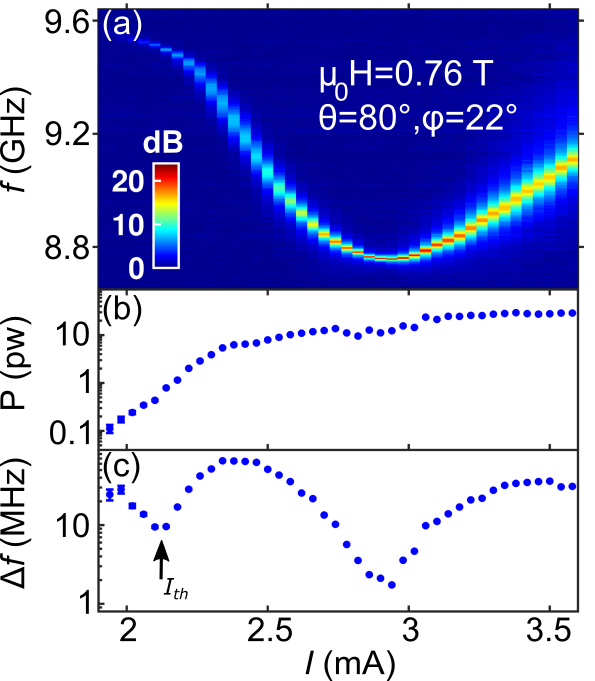}
\caption{(a) Power spectral density of a $w=$ 120 nm constriction as a function of current with a magnetic field of 0.76 T applied at an oblique angle $\theta$=80$^{\circ}$, $\varphi$=22$^\circ$. 
(b),(c) The integrated power and the line width extracted from (a).}
\label{fig:2}
\end{figure}

It should be noted that $\mathcal{N}$ also depends on the AO amplitude by itself. When the drive current, and therefore the amplitude of the AO, is small, $\mathcal{N}$ is negative for the given field and magnetization values, which results in a red shift of the AO frequency, as evident in Fig.~2(a). As the precession angle opens up with increasing drive current, the demagnetizing field decreases. This leads to an increase of the first term in Eq.~(2) and $\mathcal{N}$ becomes zero at about $I_{dc}=$ 2.9 mA. This is where the frequency red-shifting stops and a minimum can be seen in the AO frequency. Any further increase in $I_{dc}$ will make $\mathcal{N}$ positive and therefore a blue shifting frequency is observed above 2.9 mA in Fig.~2(a). 

The linewidth is also affected by the change in the nonlinearity of the system \cite{kim2008prl1,slavin2009nonlinear}:
\begin{equation}
    \Delta f= \Gamma_+ \frac{k_BT}{E(P)}\left[1+\left(\frac{\mathcal{N}}{\Gamma_{eff}}\right)^2\right]
\end{equation}
in which $\Gamma_+$ and $\Gamma_\mathit{{eff}}$ are the natural damping and effective nonlinear damping, respectively, 
and $E(P)$ is the AO energy, which can be assumed proportional to $P$ under the condition of the constant mode volume. Eq.~(3) explains the behavior of linewidth shown in Fig.~2(c). Minimum linewidth occurs approximately at the point where $\mathcal{N}=0$ \cite{tiberkevich2014sensitivity}. A similar trend for the linewidth and frequency is observed in all the other SHNOs with different widths except for the 50 nm SHNO, as shown in Fig.~3(a). For the 50 nm sample, the demagnetizing fields produced by the constriction edges become substantial, and Eq.~(2) is not valid anymore. The nonlinearity remains positive at all currents and the AO frequency has a blue-shift for the entire current range.
The details of how ultra small SHNOs operate is explained in the work of D\"urrenfeld \textit{et al.} \cite{Durrenfeld2017Nanosc}. \\

\begin{figure}
\centering
\includegraphics[width=8 cm]{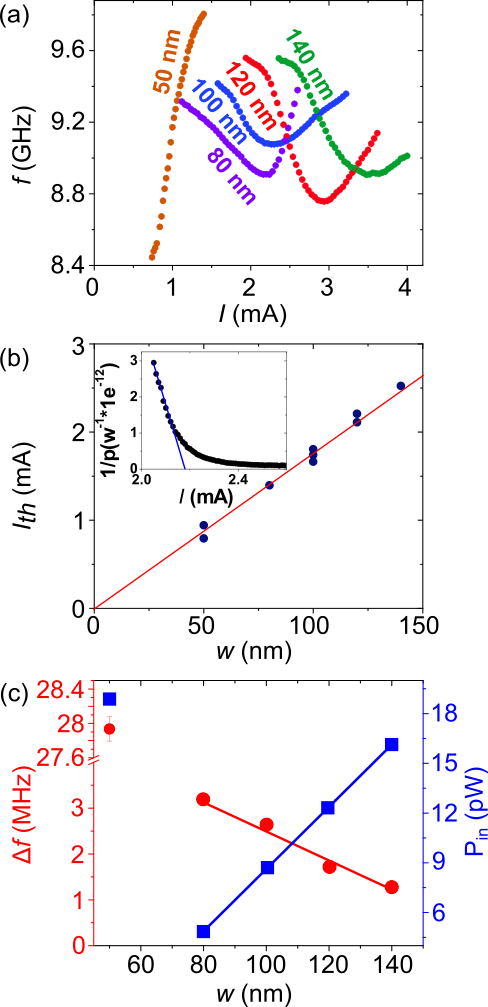}
\caption{(a) Auto-oscillation frequency vs.~drive current for all constriction widths. (b) Threshold current vs.~constriction width. Inset shows how the threshold is extracted from a linear fit of $p^{-1}$ vs.~$I$ at low currents (blue line). (c) Measured generation linewidth (red plot) and integrated power (blue plot) as functions of the constriction width; red and blue lines are linear fits.
}
\label{fig:3}
\end{figure}

The threshold current for auto-oscillations, $I_{th}$, is shown in Fig.~\ref{fig:3}(b), extracted from plots of $p^{-1}$ vs.~$I$ \cite{slavin2009ieeem} as shown in the inset.
As $I_{th}$ is found to scale linearly with the nano-constriction width, a single threshold current density, $J_{th}=$ 1.7 $\times$ 10$^{8}$ A/cm$^2$,  independent of $w$, describes all measured devices.

Microwave measurements also show that the signal characteristics of the devices improve as $w$ increases from 80 nm to 140 nm. Fig.~\ref{fig:3}(c) shows the total microwave power and the corresponding linewidth, for each device. As can be seen, the power of the devices increases linearly with $w$, while their linewidth decreases. This is consistent with the assumption, that the mode volume increases linearly with $w$, while both the spin Hall efficiency and the effective damping remain unaffected by the constriction width.
The mode with a larger volume is less prone to thermal fluctuations and therefore has a lower linewidth and at the same time higher value of the emitted power.  
However, for $w=$ 50 nm, the power of the device is larger than any other width and at the same time it has an order of magnitude higher linewidth shown in Fig.~\ref{fig:3}(c). This is likely due to the nature of the auto-oscillating mode formed in ultra-narrow devices being qualitatively different from those in larger devices. When the width of the SHNOs is large, modes first form at the edge of the constrictions and then gradually detach from the edge to move into the constriction as the drive current increases and form an interior bulk mode \cite{Dvornik2018PRA}. However, for the 50 nm device, a qualitatively different bulk mode filling up the entire constriction will form right from the beginning. This bulk mode has a higher power than the edge modes since the entire constriction is now part of the auto-oscillation but possesses a larger linewidth due to the small size of the constriction.


We have studied the operation and microwave signal properties of nano-constriction-based SHNOs in oblique magnetic fields as a function of constriction width, $w=$~50--140~nm. The frequency response 
can be explained by the nonlinearity of the system, and the threshold current is well described by a single current density, $J_{th}=$ 1.7 $\times$ 10$^{8}$ A/cm$^2$, independent of $w$. For $w=$ 80--140 nm, the quality of the generated microwave signal improves with $w$, with the output power showing a linear increase with $w$, and the linewidth showing a linear decrease.
For $w=$ 50 nm, the overall behavior is qualitatively different, with both the power and the linewidth being higher than in any other device. This is interpreted as due to a qualitatively different auto-oscillating internal bulk mode occupying the entire constriction region as previously observed in 20 nm nano-constrictions.

\begin{acknowledgments}
This work was supported by the Swedish Research Council (VR) and the Horizon 2020 research and innovation programme (ERC Advanced Grant No.~835068 "TOPSPIN").
\end{acknowledgments}
\section*{Data availability}
The data that support the findings of this study are available from the corresponding author upon reasonable request.

\end{document}